# Electrical charging of ash in Icelandic volcanic plumes

Karen L. Aplin[1,*], Isobel M. P. Houghton[2], Keri A. Nicoll[3]

1. Department of Physics, University of Oxford, Oxford, UK
2. Department of Electrical and Electronic Engineering, University of Bristol, Bristol, UK
3. Department of Meteorology, University of Reading, Reading, UK

**ABSTRACT:** The existence of volcanic lightning and alteration of the atmospheric potential gradient in the vicinity of near-vent volcanic plumes provides strong evidence for the charging of volcanic ash. More subtle electrical effects are also visible in balloon soundings of distal volcanic plumes. Near the vent, some proposed charging mechanisms are fractoemission, triboelectrification, and the so-called "dirty thunderstorm" mechanism, which is where ash and convective clouds interact electrically to enhance charging. Distant from the vent, a self-charging mechanism, probably triboelectrification, has been suggested to explain the sustained low levels of charge observed on a distal plume. Recent research by Houghton et al. (2013) linked the self-charging of volcanic ash to the properties of the particle size distribution, observing that a highly polydisperse ash distribution would charge more effectively than a monodisperse one. Natural radioactivity in some volcanic ash could also contribute to self-charging of volcanic plumes. Here we present laboratory measurements of particle size distributions, triboelectrification and radioactivity in ash samples from the Grímsvötn and Eyjafjallajökull volcanic eruptions in 2011 and 2010 respectively, and discuss the implications of our findings.

## INTRODUCTION

Volcanic ash is known to charge electrically, producing some of the most spectacular displays of lightning in nature. Factors affecting the electrical charging of ash from two different Icelandic volcanic eruptions: Eyjafjallajökull in 2010 and Grímsvötn in 2011 are presented in this paper. The Eyjafjallajökull eruption caused a well-documented flight ban over most of Europe for several days. Volcanic lightning was detected close to the vent (Bennett et al., 2010), and Harrison et al. (2010) measured the charge in the centre of the plume to be ~0.5pCm$^{-3}$ over Scotland, over 1200km from the source. This indicates a self-charging mechanism must be involved, as charge present at the vent would have decayed within a timescale of a few hundred seconds, and the distribution of charge within the plume is inconsistent with that expected from the fair weather current in the global electric circuit. The 2011 Grímsvötn eruption caused less disruption to air traffic but generated a spectacular lightning display (Arason et al, 2013). The reasons for variations in volcanic lightning and ash charging are not well understood, and have been investigated here using ash samples from the two eruptions. Ash from the 2010 Eyjafjallajökull eruption was collected at Sólheimaheiði, 22 km from the crater, and ash from the 2011 Grímsvötn eruption was collected 70 km from the crater.

Laboratory tests investigating the charge transferred to a conducting plate due to fall of volcanic ash through an insulating cylinder are described first. Gamma ray spectroscopy measurements of ash from the two volcanoes are also reported.

## MEASUREMENTS OF TRIBOELECTRIC CHARGING

A series of experiments were carried out whereby samples of volcanic ash were released to fall

Contact information: Dr Karen Aplin, Department of Physics, University of Oxford, Denys Wilkinson Building, Keble Road, Oxford OX1 3RH, UK Email: karen.aplin@physics.ox.ac.uk

vertically through a cylinder onto a screened metal plate located close to the bottom of the cylinder. The charge associated with the ash fall was measured by connecting the metal plate to an electrometer, which recorded the voltage on the plate. The apparatus consisted of a 1m long cylindrical Perspex tube, 0.25m in diameter, the top of which was sealed with an inverted funnel connected to a loading trap and shutter (made of cardboard to minimise charge generation from the apparatus). The voltage on the bottom plate was measured with a Keithley 6512 electrometer and logged to a PC via an IEEE-488 interface. Approximately 50g of ash (baked to remove adsorbed water) was loaded into the trap and the shutter released to enable ash to fall under gravity to the bottom of the tube. The tube was mounted on a support frame so that after each ash drop it was rotated to reload the ash (Krauss et al. 2003, Aplin et al. 2012).

*Results*

Eleven ash drops were obtained with the Eyjafjallajökull ash and six for the Grímsvötn ash, with typical results shown in Figure 2. The difference in polarity and magnitude of the voltage generated on the bottom plate between the two ash samples is clear.

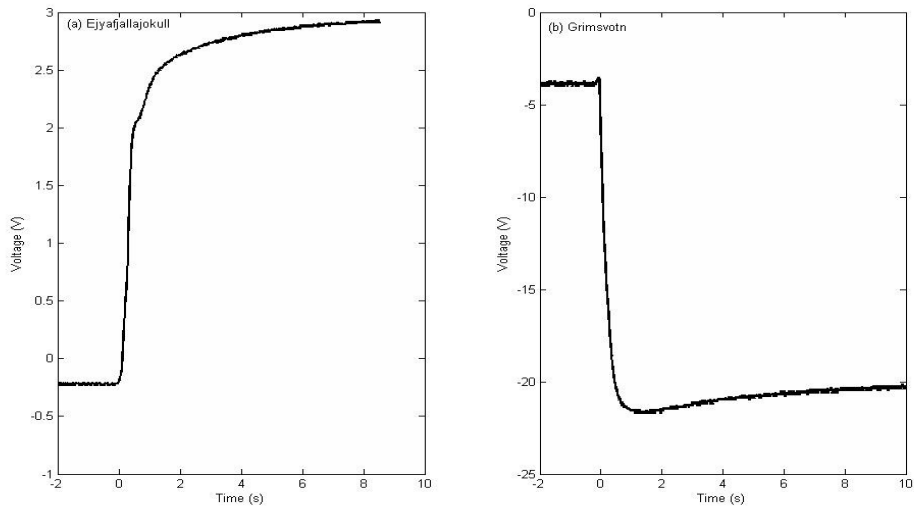

Figure 1: Typical results from ash drops. (a) shows a potential change of + 3.15V for the ash from Eyjafjallajökull, (b) the Grímsvötn ash produces a potential change of -18V.

These results indicate that the ash from Grímsvötn has a greater propensity to charge than that from Eyjafjallajökull. The sign of self-charging observed in the Eyjafjallajökull sample is consistent with the in-plume observations of Harrison et al. (2010). Further experimental work, reported in Houghton et al. (2013), found that the efficiency of triboelectric charging was related to the polydispersity of the particle size range, so that broad distributions charged more than, for example, highly bimodal distributions. This implies that all volcanic ash plumes will become charged, with implications for detection, plume lifetime and the associated aircraft hazards (Houghton et al., 2013).

The reasons why the Grímsvötn ash charged more effectively, and negatively, than Eyjafjallajökull ash are not well understood. Part of the explanation may be related to the size distribution of the ash particles, shown in figure 2.

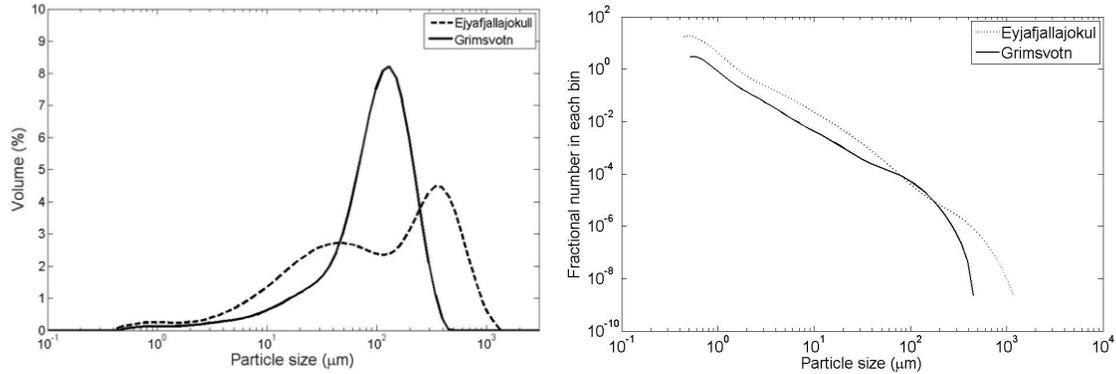

Figure 2: Size distributions of the ash samples used. (left) Volumetric size distribution, measured with a Malvern Mastersizer 2000. (right) Number size distribution, calculated from the volumetric size distribution assuming spherical particles and a packing fraction of 0.5 for Eyjafjallajökull and 0.2 for Grímsvötn.

Theoretical work (Lacks and Levandovksy, 2007) explained empirical observations that small particles charge negatively and larger ones positively through redistribution of surface electrons between different particles. Houghton et al. (2013) showed that the span (an index of the polydispersity of a particle distribution) controlled the triboelectric charging for Grímsvötn ash. The span for the Grímsvötn sample was 1.919, whereas the Eyjafjallajökull sample has a span of 1.987, indicating that between two different ash samples, span alone is not adequate to explain the observed differences in charging. Microscopic analysis of the ash particles after sieving into different size fractions indicated that the composition of each sample varied with size, so it is likely that the differing materials making up the sample triboelectrically interact, in addition to charge transfer between identical particles of different sizes.

**CHARGING FROM NATURAL RADIOACTIVITY**

As a volcanic eruption is essentially an injection of a large quantity of pulverized sub-surface rock into the atmosphere, the decay of uranium (U), thoron (Th) and potassium (K) radioisotopes naturally occurring in the rock offers another self-charging mechanism. As well as leaving a residual charge on the radioactive particle, alpha and beta particles are ionising and the ions created can attach to other particles to charge them (Clement and Harrison, 1992). Natural radioactivity levels in the ash are low compared to the surface background, since the samples did not enhance the count rate of a Geiger counter. To investigate any possible self-charging aloft from radioactivity, ash samples were placed in a Canberra 7229N gamma ray spectrometer, and data normalised for mass of the sample and against background. The results indicated a range of peaks associated with the decay of $^{238}$U, $^{232}$Th and $^{40}$K, as shown in Figure 2. Not all peaks were identified, and some peaks were indistinguishable between the two samples. The Eyjafjallajökull ash was more radioactive than Grímsvötn, and on the basis of our measurements, contained a factor of 1.6 more Th and 1.8 more U, whereas the K concentrations were indistinguishable between the two samples. No data is available on the relative U and Th concentrations in the volcanic ash, but $K_2O$ concentrations in the Eyjafjallajökull ash were a factor of 4 greater than in Grímsvötn ash (Oskarsson and Sverrisdóttir, 2011).

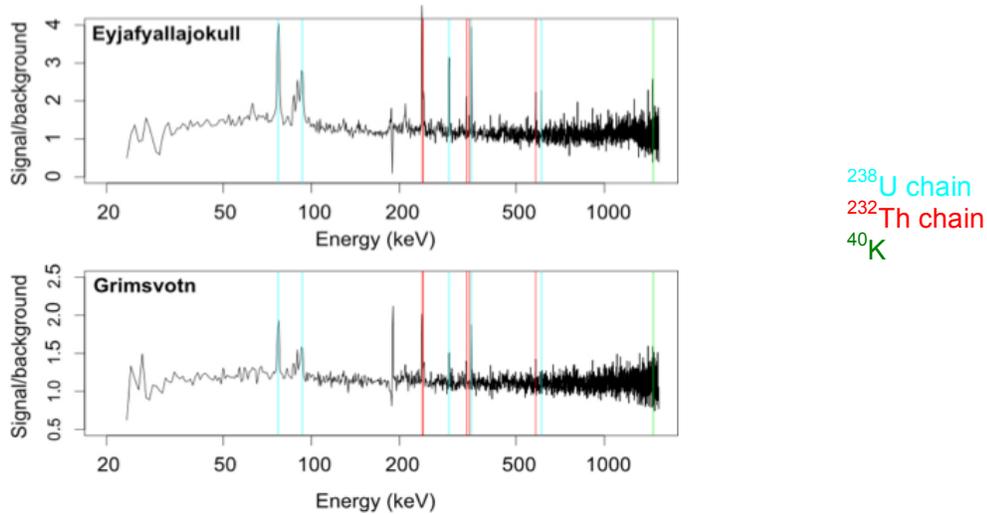

Figure 3: Gamma ray spectra, normalized against background and for mass, of ash samples from the two volcanic eruptions. Gamma peaks associated with particular natural decay chains are indicated in colour (cyan for $^{238}$U, red for $^{232}$Th and green for $^{40}$K)

As the Eyjafjallajökull ash was more radioactive than Grímsvötn, further gamma ray spectrometry measurements were carried out on the ash samples after sieving into different size fractions, to determine if the radioactivity resided in a preferred part of the size spectrum. Gamma peaks corresponding to decays of $^{214}$Bi and $^{40}$K were identified on particles > 180 μm. Alpha spectrometry measurements were also carried out using a Canberra 7401 alpha spectrometer, but no alpha decays were observed. Many other decays from the natural radioisotopes are expected to produce gamma rays with energies above the upper detection threshold of our spectrometers. However, since $^{214}$Bi is part of the $^{238}$U decay chain, it can be assumed that other isotopes in the decay chain must be present, even if they were not detected during our experiments. The average number of charges produced per ash particle can be estimated from standard information on the decay chain to be approximately 1000 on particles >180 μm, whereas Harrison et al. (2010) only observed charge on particles <10 μm. Therefore radioactivity is unlikely to have contributed to the measured self-charging of the Eyjafjallajökull plume, though it would have a proportionally greater effect nearer the vent when there are more large particles in the plume.

**CONCLUSIONS**

Laboratory measurements of the self-charging and radioactivity in volcanic ash samples from recent Icelandic eruptions have been carried out and the following conclusions can be drawn.

1. Ash from Grímsvötn becomes more triboelectrically charged than ash from Eyjafjallajökull. This is consistent with the efficient generation of lightning by Grímsvötn, although it must be noted that meteorological factors and plume height are also relevant to the formation of volcanic lightning (Arason et al., 2011).

2. When comparing ash samples from different volcanoes, the polydispersity alone is not adequate to explain the observed triboelectric charging; compositional differences should also be taken into account.

3. Ash from Eyjafjallajökull is more radioactive than ash from Grímsvötn, presumably due to different levels of U, Th and K in the underlying rock.

4. Radioactive emissions were only observed on particles exceeding 180μm, which is probably related to the ash composition. Radioactivity could therefore contribute to self-charging nearer the vent when larger particles have not been lost from the plume.

5. The results reported by Harrison et al. (2010) observing a positive charge in the distal plume are best explained by triboelectric charging, since there was no radioactivity on particles of the size seen in the plume. Additionally, laboratory experiments on triboelectric charging found that Eyjafjallajökull ash became preferentially positively charged, as observed.

Our results are also consistent with recent work indicating that charged volcanic ash can scavenge anthropogenic radioactive aerosols, which may also explain slightly enhanced radioactivity levels in the plume (Corcho Alvarado et al., 2014). Further work is needed to evaluate our measurements in the light of this data.

**ACKNOWLEDGMENTS**
Ash samples were provided by the Icelandic Met Office, and funding was obtained from the Cabot Institute, University of Bristol. Michael Humphries (Oxford University) helped with the gamma ray spectroscopy measurements, and technical support was provided by Keith Long and Peter Shrimpton (Oxford University).